# Changing Neighbors k-Secure Sum Protocol for Secure Multi-Party Computation

Rashid Sheikh, Beerendra Kumar
*SSSIST, Sehore, INDIA*
,
.

Durgesh Kumar Mishra
*Acropolis Institute of Technology and Research
Indore, INDIA .*

*Abstract*- Secure sum computation of private data inputs is an important component of Secure Multi-party Computation (SMC).In this paper we provide a protocol to compute the sum of individual data inputs with zero probability of data leakage. In our proposed protocol we break input of each party into number of segments and change the arrangement of the parties such that in each round of the computation the neighbors are changed. In this protocol it becomes impossible for semi honest parties to know the private data of some other party.

*Keywords*- Secure Multi-party Computation (SMC), Privacy, Computation Complexity, Semi honest Parties, k-Secure Sum Protocol, Information Security, Trusted Third Party (TTP).

## I. INTRODUCTION

In today's world of information technology opportunities exist for joint computation requiring privacy of the inputs. These computations occur between parties which may not have trust in one another. In literature this subject is called Secure Multi-party Computation (SMC). It is aimed at privacy of individual inputs and the correctness of the result. Formally in SMC the parties $P_1, P_2, ..., P_n$ want to compute some common function $f(x_1, x_2, ..., x_n)$ of inputs $x_1, x_2, ..., x_n$ such a party can know only its own input $x_i$ and the value of the function $f$. The SMC problems use two computation models; ideal model and real model. In ideal model there exists a Trusted Third Party (TTP) which accepts inputs from all the parties, evaluates the common function. In real model the parties agree on some protocol which allows all the parties to evaluate the function. For example if two banks cooperatively want to know details about some customer but no bank is willing to disclose the details of the customer to other bank due to privacy of customer or policy of the bank. In such situations the SMC solutions are important. The best example of SMC is the secure sum computation where all the cooperating parties want to compute the sum of their individual data inputs while preserving confidentiality of inputs [10]. The secure sum protocol proposed by Clifton *et al.* in [10] uses randomization method for computing the sum. In this protocol two adjacent parties to a middle party can collude maliciously to know the data of a middle party. We proposed new protocols in [11] where the probability of data leakage has been reduced by segmenting the data block into a fixed number of segments.

In this paper we propose a novel secure sum computation protocol with zero probability of data leakage. In this protocol we change the position of the parties so that the neighbors are changed in each round of the computation. This protocol which is an extension of our previous protocol we call as *ck-Secure Sum Protocol*.

## 2. BACKGROUND

The subject of SMC began in 1982 when Yao proposed his millionaire's problem in which two millionaires wanted to know who was richer without revealing individual wealth to each other [1]. The solution provided was for semi honest. The semi honest parties follow the protocol but also try to know some other information. The concept was extended to multi-party computation [2]. Goldreich *et al.* also used circuit evaluation protocols for secure computation. Some real life applications of SMC emerged like Private Information Retrieval (PIR) [3, 4], Privacy-preserving data mining [5, 6], Privacy-preserving geometric computation [7], Privacy-preserving scientific computation [8], Privacy-preserving statistical analysis [9] etc. An excellent review of SMC is provided by Du *et al.* in [12] where they developed a framework for SMC problem discovery and transformation of normal problem to SMC problem. A review of SMC problems with a focus on telecommunication systems is provided by Oleshchuk *et al.* in [13]. Anonymity enabled SMC was proposed by Mishra *et al.* in [14] where the identities of the parties are hidden for achieving privacy.

In this paper the protocol is motivated by the work of Clifton *et al.* [10] where they proposed a toolkit of components for solution to SMC problems. They proposed that one of components of the toolkit for SMC is the secure sum computation. Secure sum computation is used in many distributed data mining applications where many geographically distributed sites compute sum of values from individual sites. The secure sum protocol proposed in [10] used random numbers for privacy of individual data inputs. In this scheme any two parties $P_{i-1}$ and $P_{i+1}$ can collude to know the secret data of party $P_i$ by performing only one computation. We proposed *k-Secure Sum Protocol* and *Extended k-Secure Sum Protocol* in [11] where the probability of data leakage is significantly reduced by breaking the data block of individual party in number of segments. The probability of data leakage decreases as the number of segments in a data block is increased.

As per our survey no secure sum protocol is available in the literature with zero probability of data leakage when two neighbors collude. In this paper we proposed zero probability protocol for secure sum computation namely *ck-Secure Sum Protocol* in which neighbors are changed in each round of computation.

### 3. PROPOSED ARCHITECTURE AND THE PROTOCOL DESCRIPTION

The initial architecture of the protocol is shown in fig 1 where parties are arranged in a ring. Each party breaks the data block into $k$ segments which is equal to $n$-1. For example in fig 1 four parties break their data block into three segments. Initially the parties are arranged sequentially as $P_1, P_2, ..., P_n$. In the next round of the computation $P_2$ exchanges its position with $P_3$ and in subsequent rounds $P_2$ exchanges its position with $P_4$ and so on until $P_n$ is reached.

### 3.1 INFORMAL DESCRIPTION OF CK-SECURE SUM PROTOCOL

We observed in secure sum protocol [10] and k-secure sum protocol [11] that a middle party can be hacked by two neighbor parties with some probability. The motivation for *ck-Secure Sum Protocol* is that we change the neighbors in each round of segment computation. Thus it is guaranteed that no two semi honest parties can learn all the data segments of a victim party. In this protocol also each party breaks the data block into $k = n-1$ segments where $n$ is the number of parties involved in the cooperative sum computation. We propose $P_1$ to be the protocol initiator. The position of the protocol initiator is kept fixed in each round of computation. For the first round of the computation parties are arranged in a serial fashion as $P_1, P_2, ..., P_n$. The protocol initiator starts computation using k-secure sum protocol to get the sum of first segment of each party. Before second round of computation starts $P_2$ exchanges its position with $P_3$. In next round of the computation $P_2$ exchanges its position with $P_4$ and so on until $P_2$ exchanges its position with $P_n$. Generalizing the method we can say that in $i^{th}$ round of the computation $P_2$ exchanges its position with $P_{i+1}$ until $P_n$ is reached. In each round of computation segments are added using k-secure sum protocol [11] and the partial sum is passed to the next party until all the segments are added and the sum is announced by the protocol initiator party. Snapshots for a four-party case are shown in fig 2

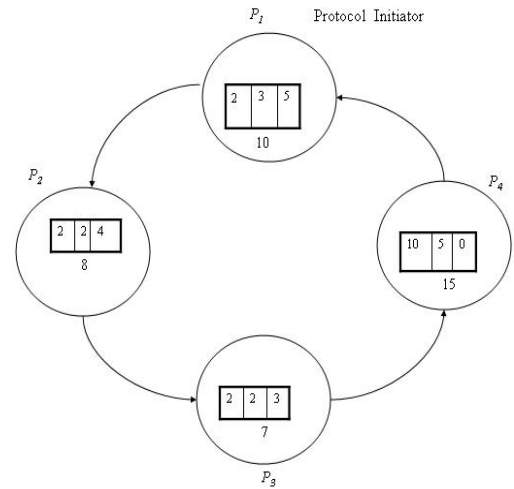

Figure 1: Initial architecture of *ck-Secure Sum Protocol*

### 3.2 FORMAL DESCRIPTION OF CK-SECURE SUM PROTOCOL

The *ck-Secure Sum Protocol* is an extension of *k-Secure Sum Protocol* [11] and is based on changing neighbors in each round of segment computation. The party $P_1$ is selected as the protocol initiator party which starts the computation by sending the first data segment. The party traverses towards $P_n$ in each round of the computation. The number of parties for this protocol must be four or more. When all the rounds of segment summation is completed the sum is announced by the protocol initiator party

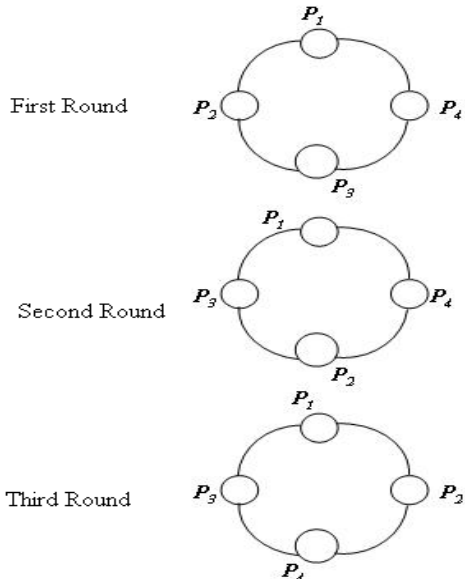

Figure 2: Snapshots of *ck-Secure Sum Protocol* for four-party case.

The algorithm: *ck-Secure Sum*
1. Define $P_1, P_2, ..., P_n$ as $n$ parties where $n >= 4$.
2. Assume these parties have secret inputs $x_1, x_2, ..., x_n$.
3. Each party $P_i$ breaks its data $x_i$ into $k = n-1$ segments $d_{i1}, d_{i2},..., d_{ik}$ such that $\sum d_{ij} = x_i$ for $j = 1$ to $k$.
4. Arrange parties in a ring as $P_1, P_2, ..., P_n$ and select $P_1$ as the protocol initiator.
5. Assume $rc = k$ and $S_{ij} = 0$. /* $S_{ij}$ is partial sum and $rc$ is round counter*/
6. While $rc != 0$
   begin
    for $j = 1$ to $k$
     begin
      for $i = 1$ to $n$
       begin
         starting from $P_1$ each party computes cumulative sum $S_{ij}$ of its next segment and thereceived sum from its neighbor and sends to the next party in the ring
       end
      $P_2$ exchanges its position with $P_{(j+2) \mod n}$
     end
    $rc = rc - 1$
   end
7. Party $P_1$ announces the result as $S_{ij}$.
8. End of algorithm.

### 3.3 PERFORMANCE ANALYSIS OF CK-SECURE SUM PROTOCOL

In this protocol each data segment is secret of the party and chosen with its own way. If two neighbor parties collude they can know only one segment in one round of the computation. The protocol guarantees that a party will not have same two neighbors in all the rounds of the computations. The neighbors are changed at least once during secure sum computation. Thus any two neighbors to a middle party cannot know all the segments of a party. The semi honest parties cannot learn more information than the result. Thus the probability of data leakage by two colluder parties to a middle party is zero. Number of rounds of computation is $n-1$ and the number of exchanges between parties is $n-2$. The only drawback of this scheme is that the topology of the computational network changes in each round of the computation. The communication and computation complexity both are $O(n^2)$.

### 3.4 CONCLUSION AND FUTURE SCOPE

Secure sum computation is an important element of toolkit for SMC solution. Protocols are needed for secure sum computation with greater security to individual data. The protocol *ck-Secure Sum Protocol* changes neighbors in each round of computation. Our proposed protocol provides zero probability of data leakage by two colluding parties when they want to attack data of a middle party. This is an appreciable improvement over previous protocols available in the literature. Efforts can be made to reduce the computation and the communication complexity preserving the property of zero hacking.

Authors Profile
*Dr. Durgesh Kumar Mishra*
Ph - +91 9826047547, +91-731-4730038
Email: durgeshmishra@ieee.org

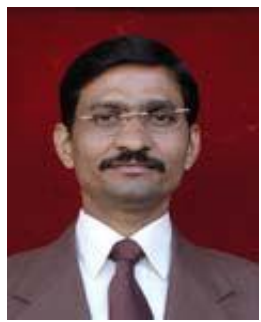

Dr. Durgesh Kumar Mishra has received M.Tech. in Computer Science from DAVV, Indore in 1994 and PhD in Computer Engineering in 2008. Presently he is working as Professor (CSE) and Dean (R&D) in Acropolis Institute of Technology & Research, Indore, MP, India. He is having around 20 Yrs of teaching experience and more than 5 Yrs of research experience. He has completed his research work with Dr. M. Chandwani, Director, IET-DAVV Indore, MP, India on Secure Multi-Party Computation. He has published more than 65 papers in refereed International/National Journals and Conferences including IEEE and ACM He is a senior member of IEEE and Secretary of IEEE MP-Subsection under the Bombay Section, India. Dr. Mishra has delivered tutorials in IEEE International conferences in India as well as other countries. He is the programme committee member of several International conferences. He visited and delivered invited talks in Taiwan, Bangladesh, USA, UK etc. on Secure Multi-Party Computation of Information Security. He is an author of one book. He is reviewer of three international journals of information security. He is Chief Editor of *Journal of Technology and Engineering Sciences*. He has been a consultant to industries and Government organization like Sales tax and Labor Department of Government of Madhya Pradesh, India.

*Rashid Sheikh*
Ph. +91 9826024087
Email: rashidsheikhmrsc@yahoo.com

*Beerendra Kumar*
Ph. +91 9770435336
Email: beerucsit@gmail.com

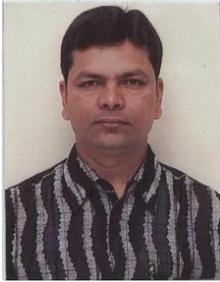

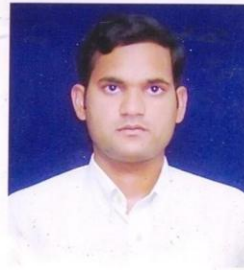

Rashid sheikh has received his Bachelor of Engineering degree in Electronics and TelecommunicationEngineering from Shri Govindram Seksaria Institute of Technology and Science, Indore, M.P., India in 1994. He has 15 years of teaching experience. His subjects of interest include Computer Architecture, Computer Networking, Electrical Circuit analysis, Digital Computer Electronics, Operating Systems and Assembly Language Programming. Presently he is pursuing M. Tech. (Computer Science and Engineering) at SSSIST, Sehore, M.P., India. He has published four research papers in National Conferences and one research paper in international journal. His research areas are Secure Multiparty Computation and MANET. He is the author of ten books on Computer Organization and Architecture.

Beerendra Kumar has received B.Tech. (Bachelor of Technology) degree in Computer Science and Information Technology from Institute of Engineering and Technology, Rohilkhand University, Bareilly (U.P), India in 2006. He has completed his M.Tech. (Master of Technology) in Computer Science from SCS, Devi Ahilya University, Indore, India in 2008.He has two years of teaching experience. His subjects of interest include Computer Networking, Theory of Computer Science, Data Mining, Operating Systems and Analysis & Design of Algorithms. He has published three research papers in national conferences and one research paper in international journal. His research areas are Computer Networks, Data Mining, Secure Multiparty Computations and Neural Networks.